\documentclass[aps,pra,twocolumn,showpacs,floatfix,nofootinbib,superscriptaddress]{revtex4}
\usepackage{amssymb,graphics,graphicx,times,color}

\input{tcilatex}
\begin{document}

\title{Non-Markovian Quantum Jumps in Excitonic Energy Transfer}
\author{Patrick Rebentrost}
\affiliation{Department of Chemistry and Chemical Biology, Harvard University, 12 Oxford
St., Cambridge, MA 02138}
\author{Rupak Chakraborty}
\affiliation{Department of Chemistry and Chemical Biology, Harvard University, 12 Oxford
St., Cambridge, MA 02138}
\author{Al\'an Aspuru-Guzik}
\affiliation{Department of Chemistry and Chemical Biology, Harvard University, 12 Oxford
St., Cambridge, MA 02138}


\begin{abstract}
We utilize the novel non-Markovian quantum jump (NMQJ) approach to
stochastically simulate exciton dynamics derived from a time-convolutionless
master equation. For relevant parameters and time scales, the
time-dependent, oscillatory decoherence rates can have negative regions, a
signature of non-Markovian behavior and of the revival of coherences.
This can lead to non-Markovian population beatings for a dimer system at room
temperature.
We show that strong exciton-phonon coupling to low frequency modes can
considerably modify transport properties. We observe increased exciton
transport, which can be seen as an extension of recent environment-assisted
quantum transport (ENAQT) concepts to the non-Markovian regime.
Within the NMQJ method, the Fenna-Matthew-Olson protein is investigated as a prototype
for larger photosynthetic complexes.
\end{abstract}

\maketitle

\section{Introduction}

The initial step in photosynthesis is the excitonic transport of the energy
captured from photons to a reaction center \cite{Blankenship02}. In this
process, highly efficient transport occurs between interacting chlorophyll
molecules embedded in a solvent and/or a protein environment \cite{Fleming09}%
. The exciton transfer dynamics has been studied utilizing F\"{o}rster theory in
the limit of weak inter-molecular coupling \cite{Forster65} or Redfield
master equations in the limit of weak exciton-phonon coupling \cite{MayBook}%
. The latter approach describes the transport as dissipative dynamics for
the reduced excitonic density matrix. Master equations are developed
starting from projector operator techniques that separate relevant (system)
from less relevant (phonon) degrees of freedom. Formally exact dynamics for
the system is described by the Nakajima-Zwanzig equation with
time-convolution \cite{Nakajima58,Zwanzig60} and the time-convolutionless
(TCL) equation \cite{Hashitsume77,Shibata77,Breuer01,Pereverzev06, Jang08}.
The first is equivalent to the chronological ordering prescription (COP)
while the second corresponds to a partial ordering prescription (POP) of the
time-ordering in a system-bath cumulant expansion \cite%
{Hashitsume77,Mukamel78, Renger02}. The convolution kernel can be
transformed into the TCL kernel by including the appropriate backward
propagation for the density matrix \cite{BreuerBook,Pereverzev06}. The
time-dependent Redfield equation is derived from a second-order approximation
in the system-bath interaction Hamiltonian \cite{BreuerBook}. Further
imposing the Markov approximation leads to the standard time-independent
Redfield equation. The dynamics of the populations of the density matrix and
that of the coherences is separated when the secular approximation is
employed, in which the master equation can be cast into Lindblad form.
Recently, Palmieri \textit{et al.} developed a prescription for
reintroducing the coupling of populations and coherences with suitably
defined Lindblad operators \cite{Palmieri09}.

Non-Markovian (NM) effects can be taken into account by the
time-convolutionless approach. Other frequently used methods explicitly
include strongly coupled modes or environmental two-level systems into the
system dynamics \cite{Brito08,Rebentrost09}. The Hilbert space size and thus
numerical effort increases concomitantly. Kubo, Tanimura \textit{et al.}
developed a hierarchical treatment where auxiliary systems describe higher
order system-bath interactions \cite{Tanimura89, Tanimura91, Ishizaki09_2}.
Xu \textit{et al.} introduced an elegant filtering method for this approach
\cite{Xu05,Shi09}. While the hierarchical treatment is formally exact for Gaussian
fluctuations, the infinite set of equations is
truncated for numerical propagation. The method was recently applied to investigate long-lived
coherence in the Fenna-Matthews-Olson (FMO) protein complex \cite%
{Engel07,Ishizaki09_3}.

A Markovian master equation in Lindblad form
can be simulated by means of the Monte-Carlo wavefunction method (MCWF) \cite%
{Dalibard92}. This numerical technique relies on the property that density
matrix evolution is equivalent to an averaging of wavefunction trajectories,
each of which is interrupted by stochastic, discontinuous quantum jumps. In
this work, we employ the non-Markovian quantum jump approach (NMQJ),
recently developed by Piilo \textit{et al}. \cite{Piilo08, Piilo09}. This
method is a generalization of the MCWF to the case of non-Markovian dynamics
derived from a TCL approach. The TCL approach can lead to time-dependent,
oscillatory decoherence rates that have negative regions, a signature of NM
behavior. These negative rates are shown to lead to a reversal
of decoherence by defining appropriate quantum jumps and jump probabilities.

Compared to the explicit numerical integration of the master equation,
the NMQJ approach has interesting features in the context of exciton transfer in chromophoric networks.
First, the NMQJ approach is,
similar to the MCWF, based on the propagation of wavefunctions and thus
scales considerably better with the system size than approaches in Liouville
space. It is therefore especially suitable for simulating large chromophoric
networks of photosynthetic antenna systems. Second, positivity violation
\cite{Pereverzev06} can be efficiently detected during the simulation by a
simple criterion for the negative jump probabilities \cite{Breuer09}. Third,
the trajectory picture allows for new insights into exciton dynamics.
Quantum jumps related to negative transition rates can restore coherence and
thus can provide an additional theoretical perspective on long-lived quantum
coherence found in photosynthetic systems such as the Fenna-Matthews-Olson
complex \cite{Engel07} and the reaction center of purple bacteria \cite%
{Lee07}. Here, we apply the NMQJ approach to dimer systems and the FMO
complex. We observe population beatings arising from
recurrence effects of the NM environment. We also find that in the non-Markovian
regime transport can be enhanced compared to
purely Markovian dynamics and thereby provide an extension to the recent ENAQT concept
\cite{Mohseni08,Rebentrost09ENAQT,Plenio08}.
These effects are pronounced in
situations when the main phonon-mode frequencies are much smaller than (i.e.
\textquotedblleft off-resonant" to) a particular system transition frequency.

In Sec. 2, we develop a time-convolutionless master equation for the
dynamics of a single excitation, leading to time-dependent rates. In Sec. 3,
we discuss the spectral density and time-dependent rates in more detail and
explain the physical situation where NM effects are considerable. In Sec. 4,
we introduce the NMQJ method in the context of excitonic energy transfer.
Finally, in Secs. 5-7, we analyze dimer systems and the Fenna-Matthews-Olson
complex.

\section{Master equation}

The transport dynamics of a single excitation is described by a master
equation for the density matrix that includes coherent evolution,
relaxation, and dephasing. In this work, we are mainly interested in the
effect of slow phonon fluctuations and the memory of the bath on the excitonic
energy transport dynamics. We utilize a time-convolutionless master equation
to second order in the exciton-phonon coupling. We employ the secular
approximation and focus on non-Markovian decoherence rates. The removal of
the secular approximation requires modifications to the NMQJ
approach and is left for future work.
The validity and limitation of the Redfield approach with respect to parameters
such as environmental coupling and temperature and the with repect to the neglect of fundamental processes
such as the molecular reorganization has been discussed in detail in \cite{MayBook,Ishizaki09_1,Egorova03}.
The complete Hamiltonian for an
interacting $N$-chromophoric system in the single exciton manifold and
including the phonon part is given by $H=H_{S}+H_{SB}+H_{B}$. The system part is
in tight-binding form:
\begin{equation}
H_{S}=\sum_{m=1}^{N}\epsilon _{m}|m\rangle \langle
m|+\sum_{n<m}^{N}V_{mn}(|m\rangle \langle n|+|n\rangle \langle m|),
\label{HamiltonianSystem}
\end{equation}%
where the Hilbert space basis states $|m\rangle $ denote the presence of an
excitation at the $m$th chromophore and $\epsilon _{m}$ are relative site
energies with respect to the chromophore with the lowest absorption energy.
The $V_{mn}$ are the interchromophoric couplings. The exciton basis $%
|M\rangle =\sum_{m}c_{m}(M)|m\rangle $ is the eigenbasis of the Hamiltonian (%
\ref{HamiltonianSystem}), $H_{S}|M\rangle =E_{M}|M\rangle $. The
exciton-phonon Hamiltonian is dominated by site energy fluctuations:%
\begin{equation}
H_{SB}=\sum_{m}A_{m}\otimes B_{m},
\end{equation}%
with the system part $A_{m}=|m\rangle \langle m|$ and the bath part $%
B_{m}=\left( \sum_{i}\hbar \omega _{i}\lambda _{i}(b_{i}+b_{i}^{\dagger
})\right) _{m}.$ Each site is separately interacting with a set of phonon
modes indicated by the subindex $m$ of the bath part. The dimensionless
coefficients $\lambda _{i}$ describe the coupling strength to each phonon
mode. The phonon Hamiltonian is $H_{B}=\sum_{i,m}\left( \hbar \omega
_{i}b_{i}^{\dagger }b_{i}\right) _{m},$ where the sum is over all phonon
modes (at each site) described by the bosonic operators $b_{i}$ and
frequencies $\omega _{i}$.
For this work, we assume that the chromophores are coupled independently
to their respective baths. Recent studies include spatial correlations into
the exciton dynamics \cite{Adolphs06,Rebentrost09_Apple,Fassioli09}.

The second order TCL master equation for the reduced system density matrix in the
interaction picture is given by \cite{BreuerBook}:
\begin{equation}  \label{MasterEquation}
\frac{d}{dt}\rho _{S}(t)=-\frac{1}{\hbar ^{2}}\int_{0}^{t}dt_{1}~\mathrm{tr}%
_{B}\left\{ [H_{SB}(t),[H_{SB}(t_{1}),\rho (t)\otimes \rho _{B}]]\right\} .
\end{equation}%
The characteristic double commutator arises from a second-order perturbation
treatment of the system-bath Hamiltonian. Note that it is assumed that the
effect of the system on the bath is small such that the total system
approximately factorizes for all times. Multiphonon processes, arising from
higher order commutators, are not taken into account. An additional
approximation in standard Redfield theory is that the phonon bath is always
in equilibrium, which neglects molecular reorganization effects. Next, we
introduce the operators $A_{m}(\omega )=\sum_{\epsilon _{M}-\epsilon
_{N}=\hbar \omega }c_{m}^{\ast }(M)c_{m}(N)|M\rangle \langle N|$, which
describe the effect of the system-bath Hamiltonian in the eigenbasis of the
system Hamiltonian, i.e. $A_{m}=\sum_{\omega }A_{m}(\omega )$, where the sum
is over all transitions in the single exciton manifold \cite{BreuerBook}.
This leads to the master equation of the form:%
\begin{eqnarray}
\frac{d}{dt}\rho _{S}(t) &=&-\frac{1}{\hbar ^{2}}\sum_{m}\sum_{\omega
,\omega ^{\prime }}[A_{m}(\omega ),[A_{m}(\omega ^{\prime }),\rho (t)]]
\label{eqTimeDependentRedfield} \\
&&\times e^{i(\omega +\omega ^{\prime })t}\int_{0}^{t}dt_{1}e^{-i\omega
^{\prime }t_{1}}S_{m}(t_{1})  \nonumber \\
&&+\frac{1}{\hbar ^{2}}\sum_{m}\sum_{\omega ,\omega ^{\prime }}[A_{m}(\omega
),\{A_{m}(\omega ^{\prime }),\rho (t)\}]  \nonumber \\
&&\times e^{i(\omega +\omega ^{\prime })t}\int_{0}^{t}dt_{1}e^{-i\omega
^{\prime }t_{1}}\frac{i}{2}\chi _{m}(t_{1}),  \nonumber
\end{eqnarray}%
with the symmetrized correlation function,
\begin{equation}
S_{m}(t)=\frac{1}{2}\mathrm{tr}_{B}\{\{B_{m}(t),B_{m}(0)\}\rho _{B}\},
\label{eqSymmetrizedCorrelator}
\end{equation}%
and the associated response function,%
\begin{equation}
\chi _{m}(t)=-i~\mathrm{tr}_{B}\{[B_{m}(t),B_{m}(0)]\rho _{B}\}.
\label{eqResponseFunction}
\end{equation}%
The quantities in Eq.~(\ref{eqSymmetrizedCorrelator}) and Eq.~(\ref%
{eqResponseFunction}) are related to the real and imaginary parts of the
bath correlator, i.e.~ $C_{m}(t)=\mathrm{tr}_{B}\{B_{m}(t)B_{m}(0)\rho
_{B}\}=S_{m}(t)+i\chi _{m}(t)/2$. The generalized time-dependent Redfield
equation (\ref{eqTimeDependentRedfield}) avoids the Markov approximation in
the sense that the upper integration limit goes to t instead of $\infty $.
One also observes the usual oscillating terms that mix population and
coherences. Next, we perform the secular approximation, essentially averaging over
these fast oscillating terms. Here, we would like to study the effect of the Markovian versus NM
decoherence rates and formulate the master equation such that the NMQJ
method can be straightforwardly applied. The secular approximation is justified in the slow decoherence regime
when $|\omega -\omega ^{\prime }|^{-1}\ll \tau _{D}$ for all transition
frequency differences, where $\tau _{D}$ is a general time scale of
decoherence.
Finally, we assume that every chromophore is embedded in an identical phonon
environment, thus the $m$ subscript for the correlator and the response function can be dropped \cite{Mohseni08}.
One arrives at the master equation in the interaction picture:%
\begin{eqnarray}
\frac{d}{dt}\rho _{S}(t) &=&\sum_{m\omega }i L(t,\omega )[A_{m}^{\dagger
}(\omega )A_{m}(\omega ),\rho (t)] \\
&&+\gamma (t,\omega )A_{m}(\omega )\rho (t)A_{m}^{\dagger }(\omega )
\nonumber \\
&&-\frac{1}{2}\gamma (t,\omega )\{A_{m}^{\dagger }(\omega )A_{m}(\omega
),\rho (t)\},  \nonumber
\end{eqnarray}%
with the time-dependent Lamb shift,
\begin{equation}
L(t,\omega )=\mathrm{Im}\{\int_{0}^{t}dt_{1}e^{-i\omega
t_{1}}C(t_{1})\},
\end{equation}%
and the time-dependent rates,
\begin{equation}
\gamma (t,\omega )=2\text{ }\mathrm{Re}\{\int_{0}^{t}dt_{1}e^{-i\omega
t_{1}}C(t_{1})\}.  \label{eqDecoherenceRate}
\end{equation}%
A transformation into the Schr\"{o}dinger picture can be readily performed,
resulting in the usual system Hamiltonian commutator term of the master
equation \cite{BreuerBook}:
\begin{eqnarray}
\frac{d}{dt}\rho _{S}(t) &=&-\frac{i}{\hbar }[H_{S}+H_{LS}(t),\rho (t)]
\label{eqNonMarkovianMasterEquation} \\
&&+\sum_{m\omega }\gamma (t,\omega )A_{m}(\omega )\rho (t)A_{m}^{\dagger
}(\omega )  \nonumber \\
&&-\sum_{m\omega }\frac{1}{2}\gamma (t,\omega )\{A_{m}^{\dagger }(\omega
)A_{m}(\omega ),\rho (t)\}.  \nonumber
\end{eqnarray}%
The Hamiltonian $H_{LS}(t)=\sum_{m\omega }L(t,\omega )A_{m}^{\dagger
}(\omega )A_{m}(\omega )$ leads to a Lamb-type renormalization of the energy
levels. In the present work, we do not consider this term since we do not expect
a qualitatively new contribution to the exciton dynamics \cite{Ishizaki09_1,Rebentrost09_Apple}.

\section{Spectral density and time-dependent rates}

\begin{figure}[tbp]
\includegraphics[scale=0.40]{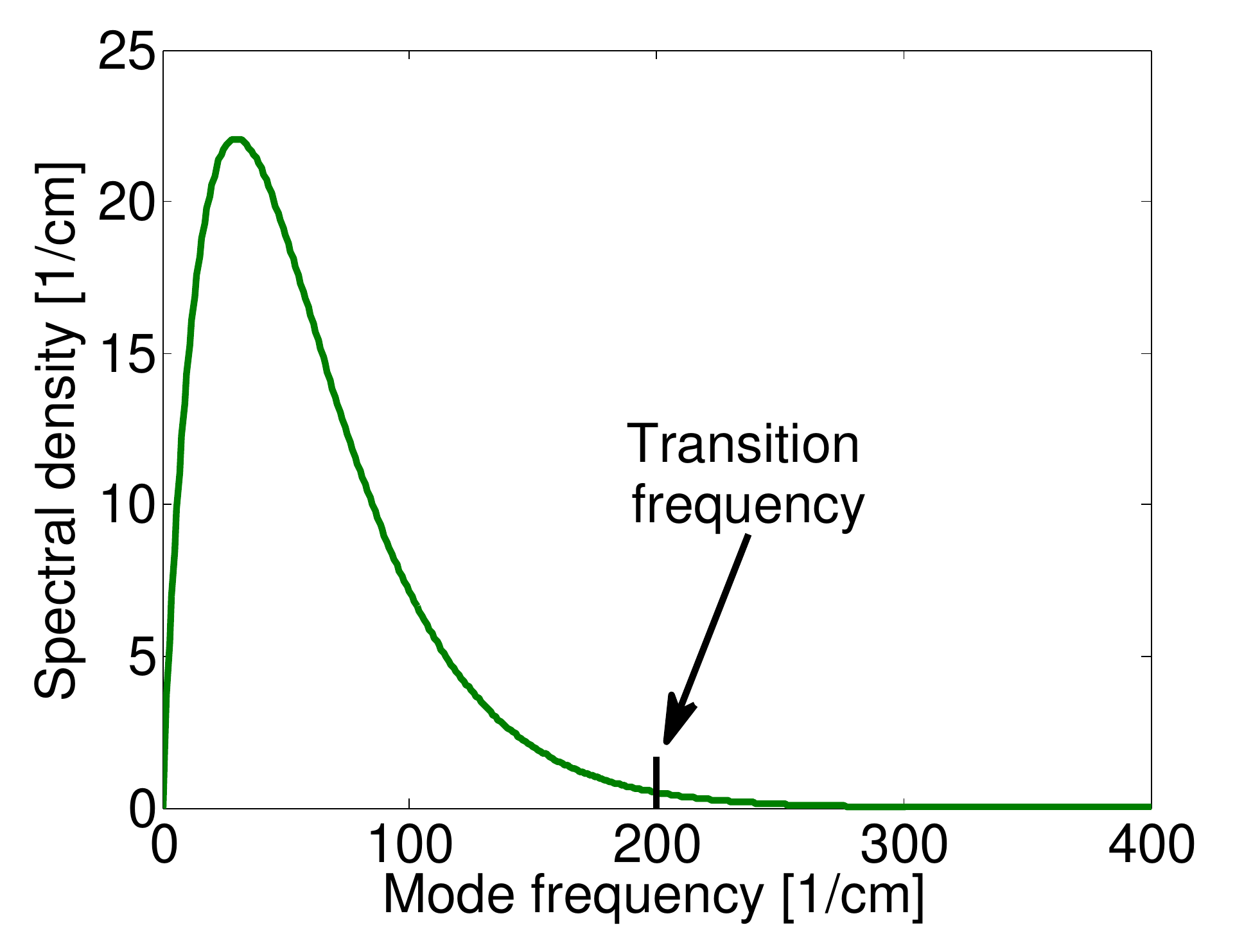} %
\includegraphics[scale=0.40]{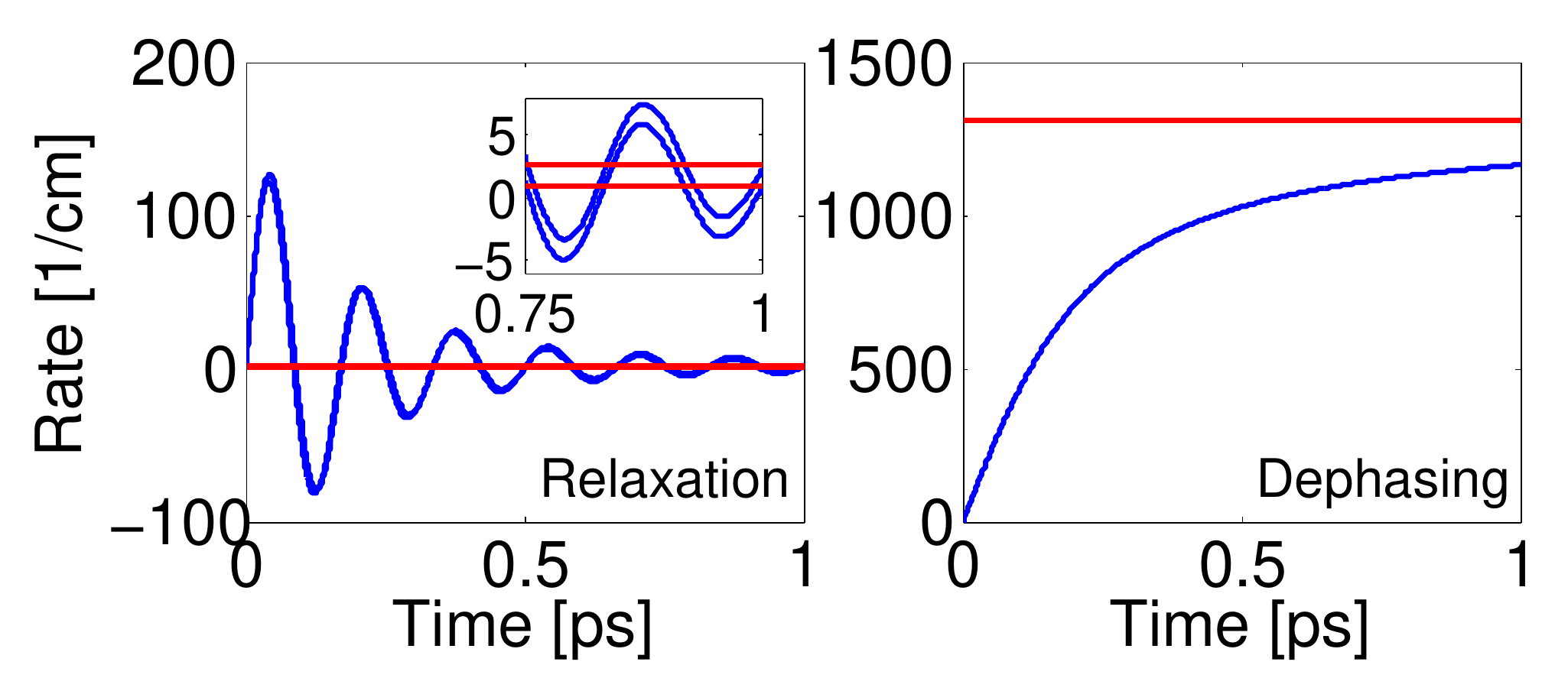}
\caption{Spectral density and resulting time-dependent decoherence rates
from Eqs.~(\ref{eqRelaxationRate}) and (\ref{eqDephasingRate}). The upper panel
shows the Ohmic spectral density with exponential cutoff for the parameters
$\lambda=30$cm$^{-1}$ and $\omega_c=30$cm$^{-1}$. In the
physical situation studied in the present work, the main strongly coupled modes are
``off-resonant" to a transition frequency (200cm$^{-1}$ in this figure).
This leads to rich behavior of the corresponding rates at relevant time
scales of around $1$ps. The relaxation rates oscillate, turn negative, and
converge to their Markovian limit (lower panel, left); the inset shows a
magnification at times just before $1$ps. The dephasing rate rises from zero
and converges to the Markovian limit (lower panel, right). }
\label{figureSpectralDensity}
\end{figure}

The main result for a Redfield master equation without Markov approximation
is the time dependence of the rates. The decoherence rates depend on the phonon
coupling strengths $\lambda _{i}$. The spectral density (units of frequency)
describes the coupling strength at a particular frequency:
\begin{equation}
J(\omega )=\sum_{i}\omega _{i}^{2}\lambda _{i}^{2}\delta (\omega -\omega
_{i}).
\end{equation}%
Assuming a continuous distribution of modes the spectral density can be modeled
with various functional forms. In molecular energy transfer often an Ohmic
spectral density with exponential or Drude-Lorentz cutoff is employed
\cite{Gilmore08, Cho05, Mohseni08, Ishizaki09_3}. In Ref.~\cite{Renger02}
non-Markovian phonon sidebands in fluorescence spectra of the B777 complex
were reproduced with a super-Ohmic spectral density. In this paper, we assume an
Ohmic spectral density with exponential cutoff:
\begin{equation}
J(\omega )=\frac{\lambda }{\hbar \omega _{c}}\omega \exp \left( -\frac{%
\omega }{\omega _{c}}\right) .  \label{eqSpectralDensityOhmic}
\end{equation}%
The relevant quantities are the cutoff $\omega _{c}$ and the reorganization
energy $\lambda$. The cutoff determines the position of the peak of the
spectral density and the reorganization energy is given by $\lambda =\hbar
\int d\omega \frac{J(\omega )}{\omega }$. In Fig.~\ref{figureSpectralDensity}
(upper panel) we show the spectral density for a particular choice of
parameters. We choose $\lambda =30$cm$^{-1}$, which is typical for
chromophores in photosynthetic systems \cite{Ishizaki09_2}, and $\omega
_{c}=30$cm$^{-1}$ which corresponds to relatively slow phonon modes. Note
that typical transition frequencies in the
single exciton manifold such as $\omega \approx 200$cm$^{-1}$ are located at the tail of the
spectral density. The resulting
Markovian relaxation rates are small. The strongly coupled,
\textquotedblleft off-resonant" modes at around $30$cm$^{-1}$ can lead to
considerable non-Markovian effects of the decoherence rates.

For any spectral density and for the bosonic bath, the time-dependent
decoherence rate is derived from Eq.~(\ref{eqDecoherenceRate}):
\begin{eqnarray}
\gamma (\omega ,t) &=&2\int_{0}^{\infty }d\tilde{\omega}J(\tilde{%
\omega})\left( n(\tilde{\omega})\frac{\sin ((\omega +\tilde{\omega})t)}{%
\omega +\tilde{\omega}}\right.  \label{eqRelaxationRate} \\
&&\left. +(n(\tilde{\omega})+1)\frac{\sin ((\omega -\tilde{\omega})t)}{%
\omega -\tilde{\omega}}\right) .  \nonumber
\end{eqnarray}
Here, $n(\omega )$ is the bosonic distribution function. In the Markovian
case $(t\rightarrow \infty )$, the spectral density is sampled only at the
frequency $\omega $, seen from the limiting behavior of the terms $\frac{1}{%
\omega \pm \tilde{\omega}}\sin ((\omega \pm \tilde{\omega})t).$ For the
dephasing rate one obtains from Eq.~(\ref{eqRelaxationRate}) in the limit
$\omega \rightarrow 0$:
\begin{equation}
\gamma^{\phi }(t)=2\int_{0}^{\infty }d\tilde{\omega}J(\tilde{\omega}%
)\coth \left( \frac{\hbar \omega }{2k_{B}T}\right) \frac{\sin (\tilde{\omega}%
t)}{\tilde{\omega}}.  \label{eqDephasingRate}
\end{equation}
In the Markovian limit, the dephasing rate becomes linearly proportional to the
temperature and the derivative of the spectral density at zero frequency
\cite{BreuerBook}. In the non-Markovian case, a greater part of the spectral
density is taken into account. This can lead to rich behavior of both relaxation
and dephasing rates. In Fig.~\ref{figureSpectralDensity} (lower panel), we
show the rates that follow from the spectral density
(\ref{eqSpectralDensityOhmic}) and the above choice of parameters. The relaxation
rates in the NM case oscillate around the Markovian rates, have positive and negative
regions, and finally converge to the Markovian rate on a time scale of 1ps.
The NM dephasing rate converges from below to the Markovian limit on a
similar time scale. This transient regime has been discussed in terms of
slippage in the initial conditions e.g.~in \cite{Suarez92}.

\section{Non-Markovian quantum jumps}

In this work, we perform a stochastic unraveling of the master equation with
the non-Markovian quantum jump approach established in Ref. \cite{Piilo08, Piilo09}.
The master equation (\ref{eqNonMarkovianMasterEquation}) is precisely in the
form required for the NMQJ method. We give a brief summary of this
technique. For every time $t$, one can separate the set of jump generators $%
A_{m}(\omega )$ into $A_{m}^{+}(\omega )$ and $A_{m}^{-}(\omega )$ depending
on the overall sign of the corresponding rate. That is for all $%
A_{m}^{+}(\omega )$ the rate is $\gamma ^{+}(\omega ,t)>0$, while for
all $A_{m}^{-}(\omega )$ the rate is $\gamma ^{-}(\omega ,t)<0$. In the
presence of only positive channels, the original MCWF method can be employed \cite{Dalibard92,Piilo08}. The
particular structure of the jump generators $A_{m}(\omega )$ (see previous
sections) leads to a relatively straightforward description of the quantum
mechanical ensemble. The density matrix at all time can be written as:
\begin{equation}
\rho (t)=\frac{N_{0}(t)}{N}|\psi _{i}(t)\rangle \langle \psi
_{i}(t)|+\sum_{M}\frac{N_{M}(t)}{N}|M\rangle \langle M|.
\end{equation}%
Here, $|\psi _{i}(t)\rangle $ is the initial state with statistical weight $%
N_{0}(t)/N.$ The exciton states $|M\rangle $ are as defined above and have a
statistical weight $N_{M}(t)/N$. Initially, $N_{0}(0)=N$ and at all times
the numbers $N_{0}(t)$ and $N_{M}(t)$ conserve probability, i.e.$N_{0}(t)+$ $%
\sum_{M}N_{M}(t)=N.$ The time evolution consists of propagation of $|\psi
_{i}(t)\rangle $ and stochastic changes of the weights $N_{0}(t)$ and $%
N_{M}(t)$. In general, one defines the effective Hamiltonian:
\begin{equation}
H_{\mathrm{eff}}=H_{S}-\frac{i\hbar }{2}\sum_{m,\omega }\gamma (t,\omega
)A_{m}^{\dagger }(\omega )A_{m}(\omega ),
\end{equation}%
where the sum is over positive and negative channels. The NMQJ method now describes the time
evolution of the ensemble $\rho (t)$ as a wavefunction evolution of the ensemble states with $H_{%
\mathrm{eff}}$ interrupted by probabilistic, discontinuous jumps
corresponding to the jump operators of all channels. Consider now a
particular ensemble member $|\psi (t)\rangle $ at time $t$ evolving for a
small time step $\delta t$. As in the MCWF, the no-jump evolution is:
\begin{equation}
|\psi (t+\delta t)\rangle =\frac{\left( 1-\frac{i\delta t}{\hbar }H_{\mathrm{%
eff}}\right) |\psi (t)\rangle }{||\left( 1-\frac{i\delta t}{\hbar }H_{%
\mathrm{eff}}\right) |\psi (t)\rangle ||}.
\end{equation}%
The positive jumps occur with probability $P_{m\omega }^{+}(t)=\delta
t~\gamma^{+}(\omega ,t)~\langle \psi (t)|A_{m}^{+\dagger }(\omega
)A_{m}^{+}(\omega )|\psi (t)\rangle $ and an ensemble member jumps according
to:
\begin{equation}
|\psi (t)\rangle \rightarrow |\psi ^{\prime }(t+\delta t)\rangle =\frac{%
A_{m}^{+}(\omega )|\psi (t)\rangle }{|||A_{m}^{+}(\omega )\psi (t)\rangle ||}%
.
\end{equation}%
The negative jumps occur from the source state $|\psi (t)\rangle $ to
target states $|\psi ^{\prime }(t)\rangle $ if the source $|\psi (t)\rangle
\ $\ has the property that it can be reached by a jump with $%
A_{m}^{-}(\omega )$ from the target state:
\begin{equation}
|\psi (t)\rangle =\frac{A_{m}^{-}(\omega )|\psi ^{\prime }(t)\rangle }{%
|||A_{m}^{-}(\omega )\psi ^{\prime }(t)\rangle ||}\rightarrow |\psi ^{\prime
}(t+\delta t)\rangle
\end{equation}%
Note that a negative jump can \textquotedblleft undo" positive jumps that
occurred earlier. The negative jump probability depends on the target state $%
|\psi ^{\prime }(t)\rangle $ and is $P_{m\omega }^{-}(t)=\frac{N^{\prime }(t)%
}{N(t)}\delta t~|\gamma^{+} (\omega ,t)|~\langle \psi ^{\prime
}(t)|A_{m}^{-\dagger }(\omega )A_{m}^{-}(\omega )|\psi ^{\prime }(t)\rangle
, $ where $N^{\prime }(t)$ is the number of ensemble members in the target
state and $N(t)$ are the number of ensemble members in the source state. A
Monte Carlo unraveling according to this prescription is shown to be
equivalent to master Eq.~(\ref{eqNonMarkovianMasterEquation}), see Ref.~\cite%
{Piilo08}. Non-Markovian quantum jumps can explicitly lead to restored
quantum coherence with this jump description that correctly handles negative
rates in the master equation.

We end this section with a note regarding positivity of the density matrix. The master equation (\ref%
{eqNonMarkovianMasterEquation}) with time-dependent rates is not guaranteed
to ensure positivity of the density matrix. However, the NMQJ method yields
a simple criterion for detecting when positivity is about to be violated based on the
negative jump probability \cite{Breuer09}. The $P_{m\omega }^{-}(t)$ is
inversely proportional to the number of ensemble members in the source state
$N(t)$. The case when $N(t)$ becomes zero and the rate is negative at the same time
is
precisely when the master equation would violate positivity. The
interpretation of this is that the environment tries to \textquotedblleft undo"
an event that never happened. Thus, based on the singularity of the negative jump
probability one can easily detect unphysical time evolution in the
algorithm. All results in this work originate from physical time evolution.

\section{Population beatings in a dimer system}

\begin{figure*}[tbp]
\includegraphics[scale=0.80]{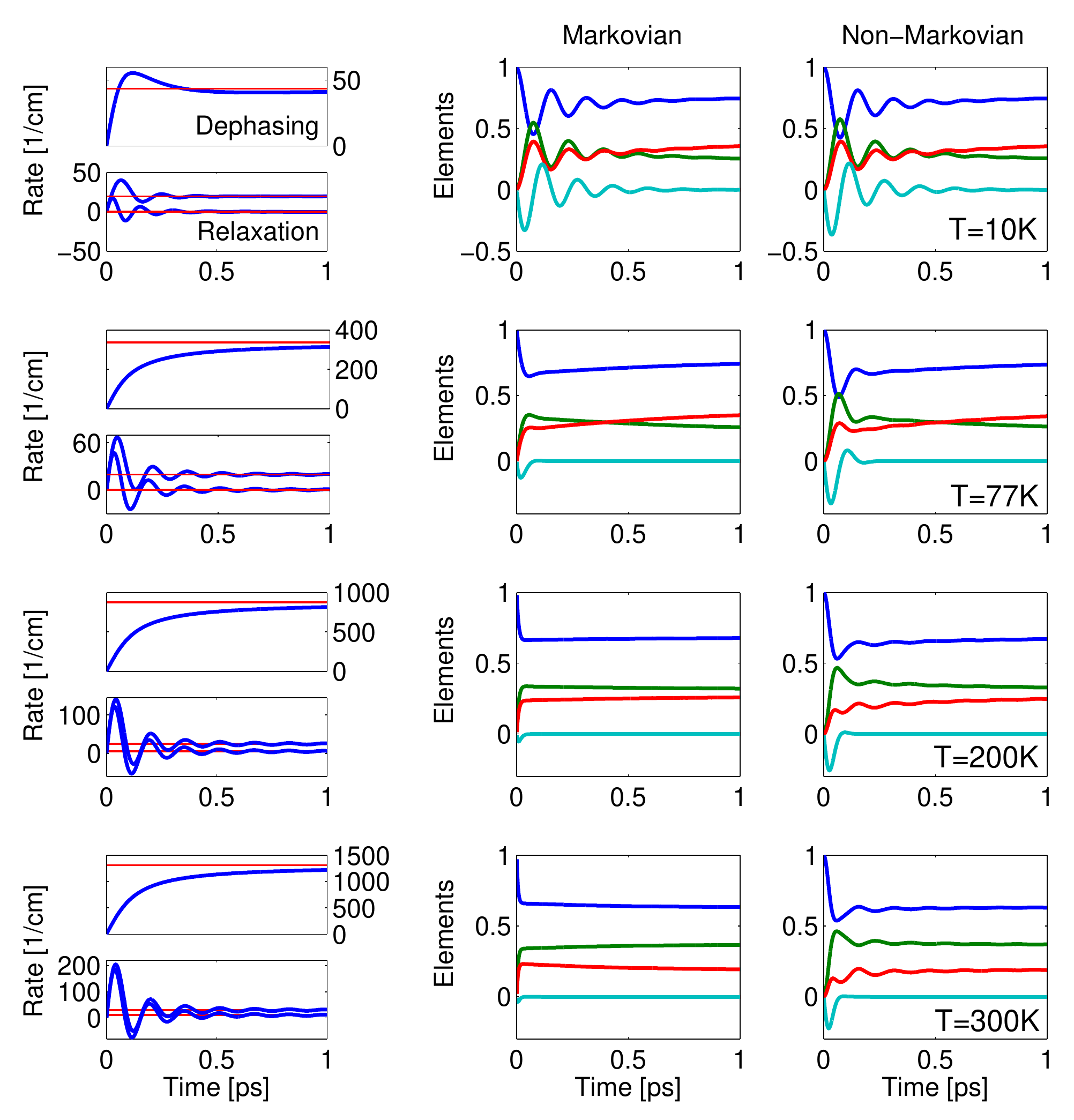}
\caption{Decoherence rates and density matrix elements for a dimer that
resembles site 1 and 2 of the Fenna-Matthews-Olson complex for Markovian and
non-Markovian evolution. The left panels show dephasing (respective upper
panel) and relaxation rate (respective lower panel) as a function of time.
The time-dependent rates (blue) converge to their respective Markovian
limits (red). Time evolution of the population elements ($\protect\rho_{11}$
blue, $\protect\rho_{22}$ green) and coherence elements (Re$\protect\rho%
_{12} $ red, Im$\protect\rho_{12}$ cyan) is displayed in the Markovian
(middle panels) and non-Markovian (right panels) case for various
temperatures. At room temperature NM population beatings are observed. }
\label{figureDimerBeatings}
\end{figure*}

In this section, we show that oscillatory non-Markovian decoherence rates can lead to
beatings of site populations that do not occur in a Markovian treatment of
the dynamics. The beatings arise from the coupling to slow modes in the
environment. We discuss a dimer system consisting of two interacting
chromophores in a structured phonon bath. The system Hamiltonian in the
single exciton manifold is:
\begin{equation}
H_{S}=\epsilon _{2}|2\rangle \langle 2|+V_{12}\left( |1\rangle \langle
2|+|2\rangle \langle 1|\right) .  \label{eqDimerHamiltonian}
\end{equation}%
The eigenenergies are E$_{2,1}=\epsilon _{2}/2\pm 1/2\sqrt{\epsilon
_{2}^{2}+4V_{12}^{2}}$ and the transition frequency is $\hbar\omega
_{21}=(E_{2}-E_{1}) =\sqrt{\epsilon _{2}^{2}+4V_{12}^{2}}$. The
respective eigenstates are $|E_{1}\rangle =c_{1}(1)|1\rangle
+c_{2}(1)|2\rangle $ and $|E_{2}\rangle =c_{1}(2)|1\rangle
+c_{2}(2)|2\rangle $, with $c_{1}(1)=-c_{2}(2)=\sin \theta $ and $%
c_{1}(2)=c_{2}(1)=\cos \theta $ with the mixing angle $\tan 2\theta
=2V_{12}/\epsilon _{2}.$ The jump generators for relaxation are $%
A_{2}(\omega )=-A_{1}(\omega )=1/2\sin 2\theta |E_{1}\rangle \langle E_{2}|$
and their transpose conjugates. The jump generators for dephasing are $%
A_{1}(0)=\sin ^{2}\theta |E_{1}\rangle \langle E_{1}|+\cos ^{2}\theta
|E_{2}\rangle \langle E_{2}|$ and $A_{2}(0)=\cos ^{2}\theta |E_{1}\rangle
\langle E_{1}|+\sin ^{2}\theta |E_{2}\rangle \langle E_{2}|.$

For the simulations we take the excitonic Hamiltonian to be of a particular
form: $V_{12}=87$cm$^{-1}$ and $\epsilon _{2}=120$cm$^{-1}$. This form is
equal to the Hamiltonian of the site 1 and 2 subsystem in the
Fenna-Matthews-Olson complex given in \cite{Ishizaki09_3}. Nevertheless, the effects
presented here hold for a large variety of dimer Hamiltonians. The initial
state is localized at site 1, i.e.~$\rho (0)=|1\rangle \langle 1|$. In Fig.~%
\ref{figureDimerBeatings}, we show the time dependence of the decoherence
rates (left panels) and the time evolution of the population and coherence
elements of the density matrix in the site basis, i.e. $\rho
_{mn}(t)=\langle m|\rho (t)|n\rangle$. We compare different temperatures
in the Markovian (middle panels) and the non-Markovian description (right panels).
The spectral density parameters are reorganization energy $\lambda =50$cm$%
^{-1}$ and cutoff $\omega _{c}=50$cm$^{-1}$. The rates (left panels in Fig.~%
\ref{figureDimerBeatings}) are similar as discussed for Fig. \ref%
{figureSpectralDensity}. The relaxation rates are generally smaller than the
pure dephasing rate. The Markovian limit is reached on a time scale on the
order of 1ps. Higher temperature leads to larger oscillations of the
decoherence rates. In the Markovian case (middle panels in Fig.~\ref%
{figureDimerBeatings}), we observe that the population and coherence
oscillations die out very fast, especially at high temperatures. This is
explained by the linear dependence of the dephasing rate on temperature.

In the non-Markovian case, the dynamics is considerably different. At low
temperatures the beatings are similar to the Markovian case; decoherence
rates are generally rather small and differences in NM versus Markovian are
not pronounced. At increasing temperature, the quantum mechanical beatings
live slightly longer in the NM case due to the smaller dephasing rate at
short times. Quantum beatings can be recognized by an oscillating imaginary
part of the coherence element of the density matrix. At large temperatures,
another type of beating arises, which is due to the oscillatory relaxation
rates. It leads to beatings of the populations matrix elements and the real
part of the coherence matrix element. These beatings can be interpreted as
recurrence of the NM environment; energy is emitted from the system into the
environment during the positive regions of the decoherence rates and
reabsorbed in the \textit{same} decoherence channel during the negative
regions.

\section{Transport in the non-Markovian regime}

\begin{figure*}[tbp]
\includegraphics[scale=0.7]{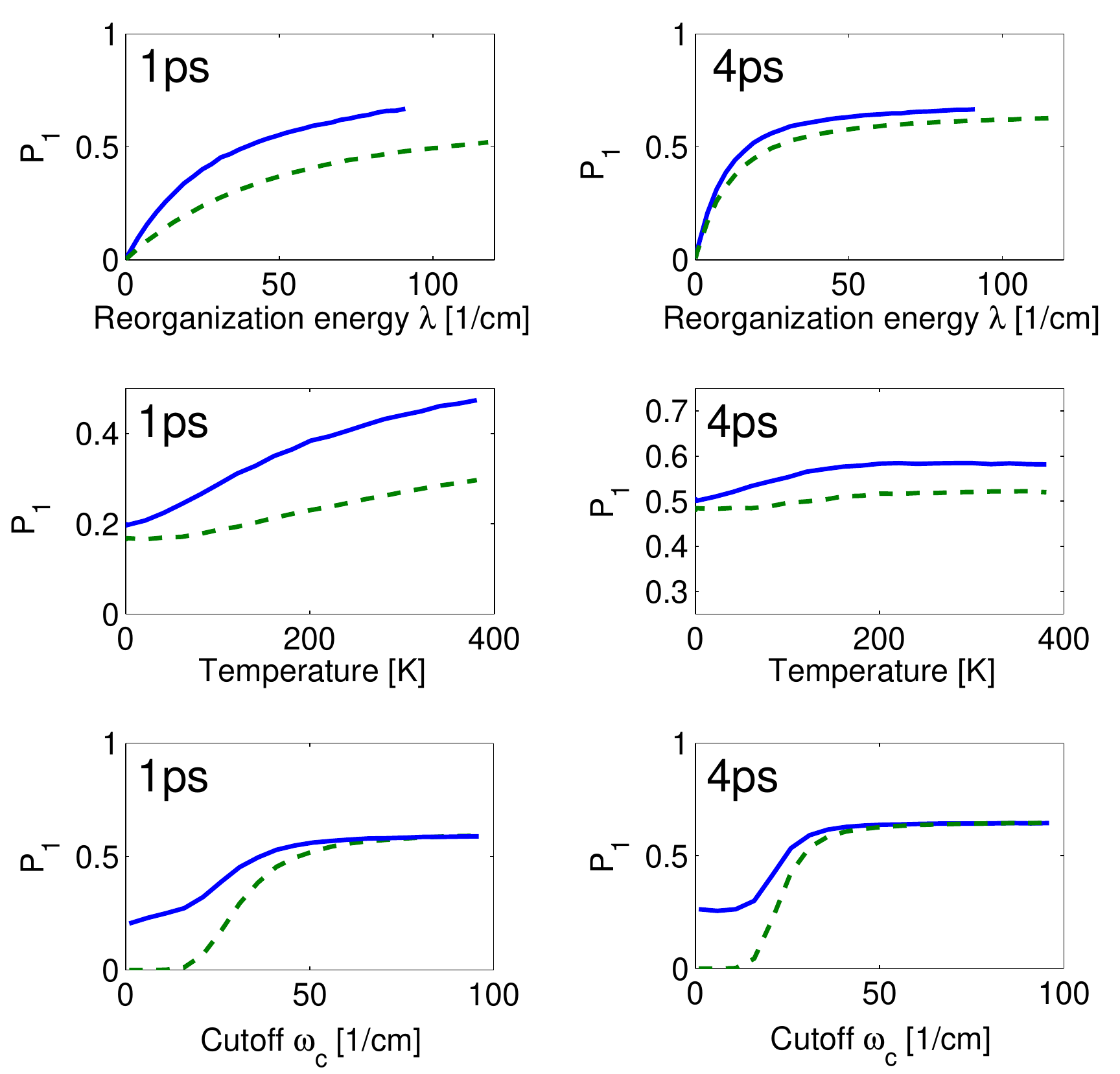}
\caption{The averaged probability measure $\bar{P}_1$ as defined in Eq.~(%
\protect\ref{eqMeasure}) as a function of the main decoherence parameters
for a dimer system. Markovian (green) and non-Markovian (blue) cases are
depicted for different temperatures and the central parameters of the
spectral density (reorganization energy $\protect\lambda$, cutoff $\protect%
\omega_c$). Two integration times for the measure, $\protect\tau=1$ps (left
panels) and $\protect\tau=4$ps (right panels), are shown, the effects being
more pronounced for the shorter time scale. The standard parameters are $%
\protect\omega_c=30$cm$^{-1}$, $\protect\lambda=30$cm$^{-1}$, and room
temperature $T=300$K. }
\label{figureDimerScans}
\end{figure*}

In this section, we focus on transport properties in the non-Markovian regime.
The general behavior
of the exciton transport as a function of different parameters,
contributions of various physical processes, and robustness of a
chromophoric network can be investigated by theoretical measures such as
the energy transfer efficiency and the transfer time \cite%
{Castro07,Rebentrost09_Apple}. Trapping sites model the reaction centers
where charge separation and energy storage occurs in the photosynthetic
system, neglecting further chemical detail. In this paper, we utilize a
simpler measure to elucidate energy transport. We define the integrated
probability of a particular excitonic state up to a certain time $\tau $,
the only free parameter. An explicit introduction of trapping sites and
additional free parameters is not required. Formally, the measure is given
by:
\begin{equation}
\bar{P}_{M}=\frac{1}{\tau }\int_{0}^{\tau }dt\text{ }\langle M|\rho
(t)|M\rangle ,  \label{eqMeasure}
\end{equation}%
where $\tau $ is the total integration time and $|M\rangle $ is a particular
exciton state given by the problem at hand. For this measure we choose an
exciton state $|M\rangle $ and focus on relaxation dynamics in the exciton
basis. For example, in the Fenna-Matthews-Olson complex the exciton with the
lowest energy, localized at site 3 and 4, would be an appropriate choice. A
site-basis definition is straightforward within the NMQJ method.

We analyze the transport properties of master equation (\ref%
{eqNonMarkovianMasterEquation}) and the NMQJ unraveling and compare to the
Markovian regime. We choose the Hamiltonian parameters in Eq. (\ref%
{eqDimerHamiltonian}) as $V_{12}=50$cm$^{-1}$ and $\epsilon _{2}=2V_{12}$.
We assume that the system is initially in the energetically higher
eigenstate $|E_{2}\rangle $ and investigate relaxation to the lower lying
eigenstate $|E_{1}\rangle $. We quantify the transport by the integrated
probability of Eq.~(\ref{eqMeasure}) using $|E_{1}\rangle $, i.e. $\bar{P}_1$%
. In Fig.~%
\ref{figureDimerScans}, we show the dependence of the transport on the essential
parameters of the spectral density (reorganization energy $\lambda $ and
cutoff $\omega _{c}$) and the temperature. If not explored as variables, the default parameters are $\lambda =30$cm$%
^{-1}$, $\omega _{c}=30$cm$^{-1}$, and room temperature ($T=300$K).
We investigate two integration
times, $\tau =1$ps and $\tau =4$ps. The results are more pronounced
for the shorter time. Shorter time scales are more relevant for smaller
photosynthetic complexes such as the Fenna-Matthews-Olson complex \cite%
{Ishizaki09_3}. As a result, we observe that transport can be enhanced
in the NM situation where the rates are gives by Eq.~(\ref{eqRelaxationRate}).
This study can be seen as an extension of ENAQT concepts to the non-Markovian regime.

In the upper panels of Fig.~\ref{figureDimerScans}, the dependence of $\bar{P%
}_1$ as a function of the reorganization energy $\lambda$ is investigated.
Relaxation and dephasing rates are scaled linearly with $\lambda$. In general,
the probability $\bar{P}_1$
increases as a function of the reorganization energy. When relaxation
rates are larger, thermal equilibration of the exciton populations is
faster. At $\lambda =30$cm$^{-1}$ and $\tau =1$ps the non-Markovian probability is $\bar{P}%
_1$ $=0.44$ while the Markovian probability is $\bar{P}_1=0.27$, a substantial difference
for this rather small system. The improvement can be rationalized by the
fact that the initially large NM relaxation rates lead to fast
equilibration. The negative regions of the rate partially \textquotedblleft
undo" the positive region but the integrated population of the target
exciton is overall larger than in the Markovian case. For reorganization
energies beyond  $\approx$90cm$^{-1}$ we observe positivity violating time
evolution identified with the criterion discussed earlier \cite{Breuer09}.

The middle panels of Fig.~\ref{figureDimerScans} show the dependence of $\bar{P}_1$ on
the temperature. In the present case of energy transfer from a
high to low exciton state, temperature can have an assisting effect for short
times and for both Markovian and NM treatment \cite{Mohseni08}. For example, see the graphs for $\tau=1$ps. Increased thermal population of the
phonon modes can lead to increased stimulated emission of exciton energy
into the phonon bath and thus transport towards the lower exciton state.
This effect becomes weaker for longer times, see the graphs for $\tau=4$ps.
Absorption of energy from the phonon bath comes into play which transports
the excitation from the lower exciton state back to the higher one. The
temperature is more significant in the NM regime since the Bose functions in
the rate integral are sampled at all frequencies instead of only at $\omega
_{21}$.

In the lower panels of Fig.~\ref{figureDimerScans}, the averaged probability
$\bar{P}_{1}$ is shown as a function of the cutoff $\omega _{c}$ of the
spectral density. For both NM and Markovian cases, a larger cutoff and thus
a stronger coupling of modes that are resonant with the transition frequency
leads to increased transport. The differences in terms of transport of both
cases vanish when $\omega _{c}\approx 60$cm$^{-1}$; resonant modes dominate
the relaxation dynamics. However, in the presence of only slow modes, the NM
treatment shows substantially larger transfer probabilities. For $\tau =1$ps and $\omega
_{c}=20$cm$^{-1}$, we obtain $\bar{P}_{1}=0.31$ in the NM case and $\bar{P}%
_{1}=0.06$ in the Markovian case. This improvement is due to sampling of the
a broader range of the spectral density in Eq.~(\ref{eqRelaxationRate}).
The physical interpretation is that the NM treatment
allows the system to temporarily access energy non-conserving phonons for
quantum jumps that would be inaccessible otherwise.

\section{Fenna-Matthews-Olson complex}

\begin{figure*}[tbp]
\includegraphics[height=0.24\textheight]{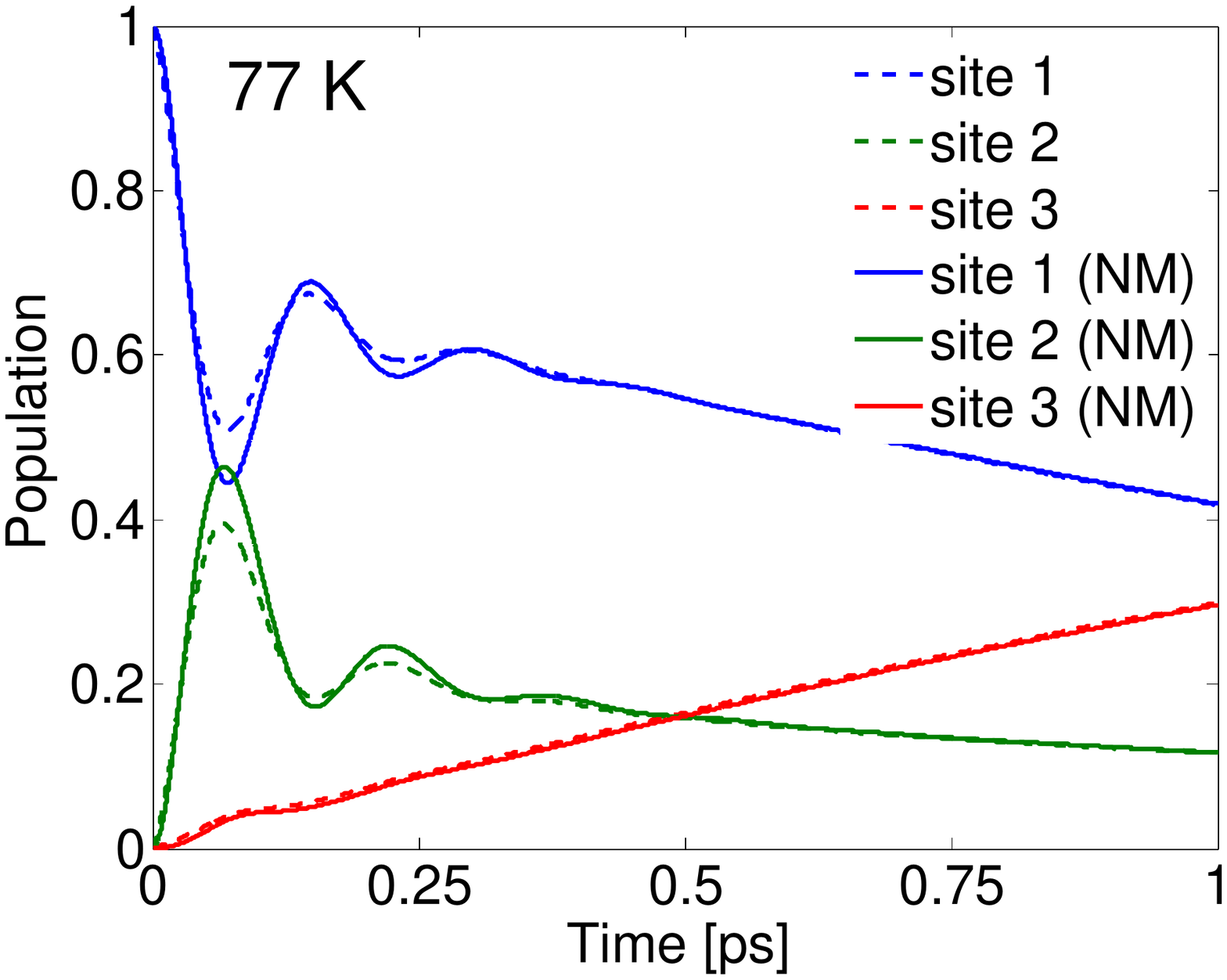} %
\includegraphics[height=0.24\textheight]{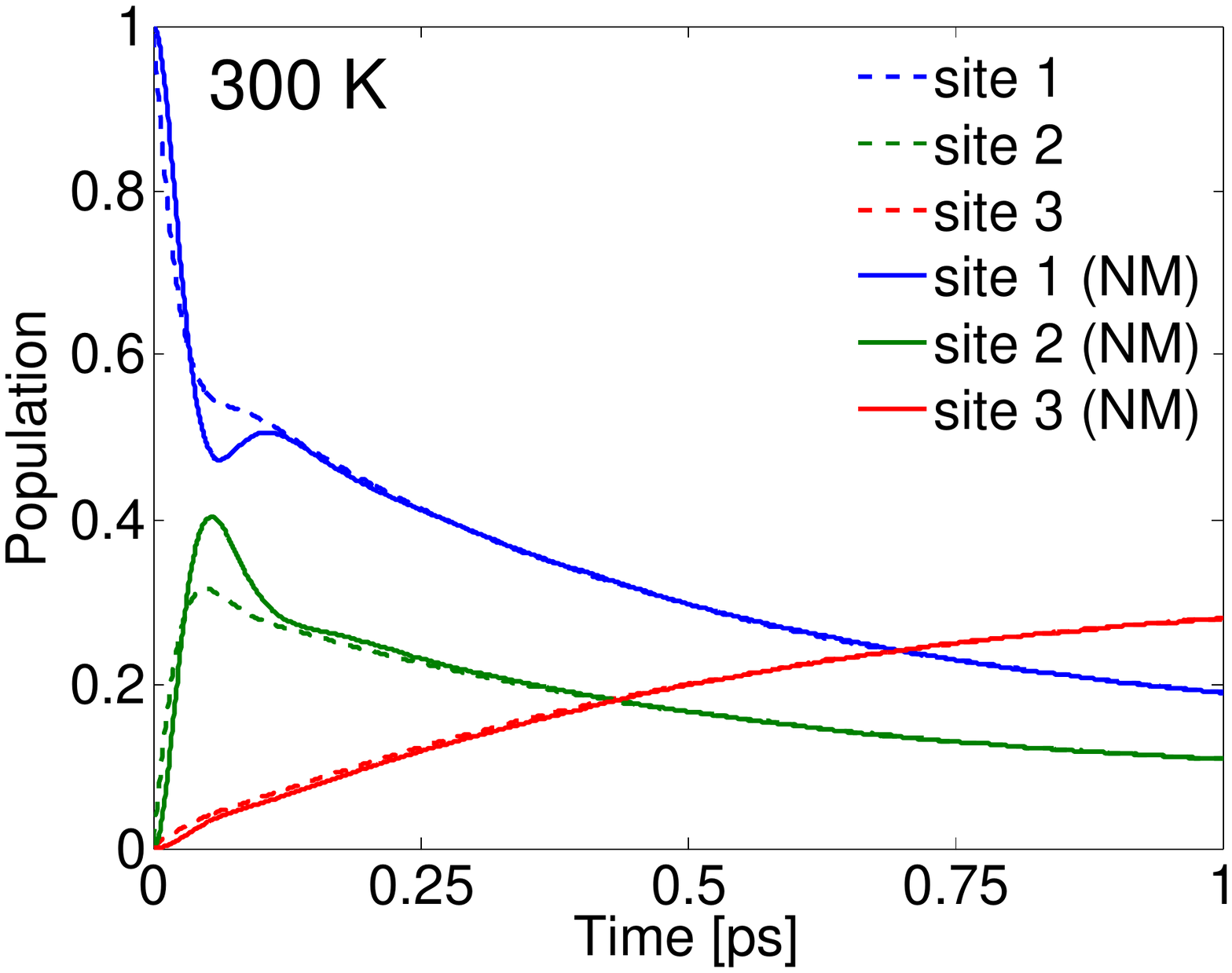} %
\includegraphics[height=0.24\textheight]{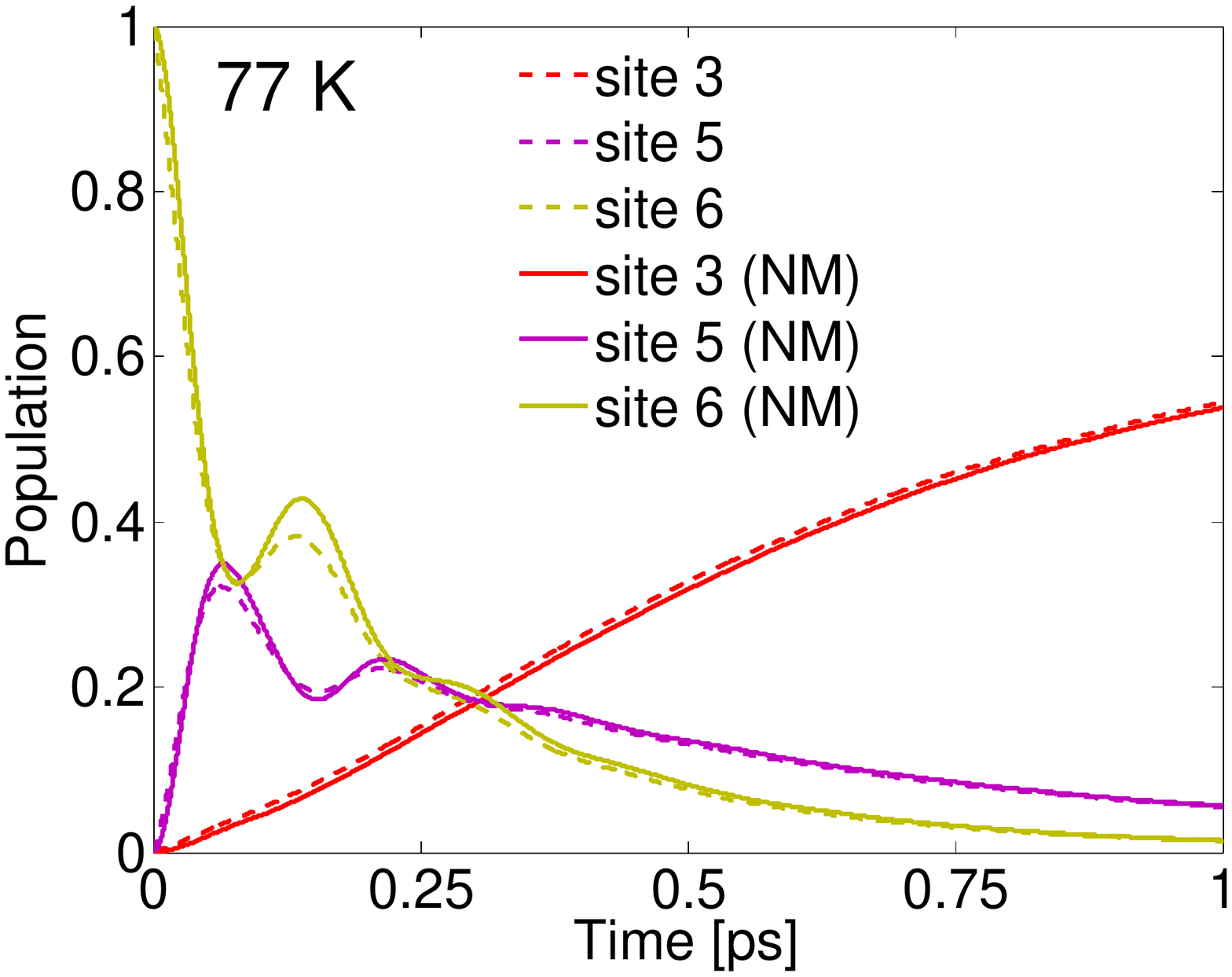} %
\includegraphics[height=0.24\textheight]{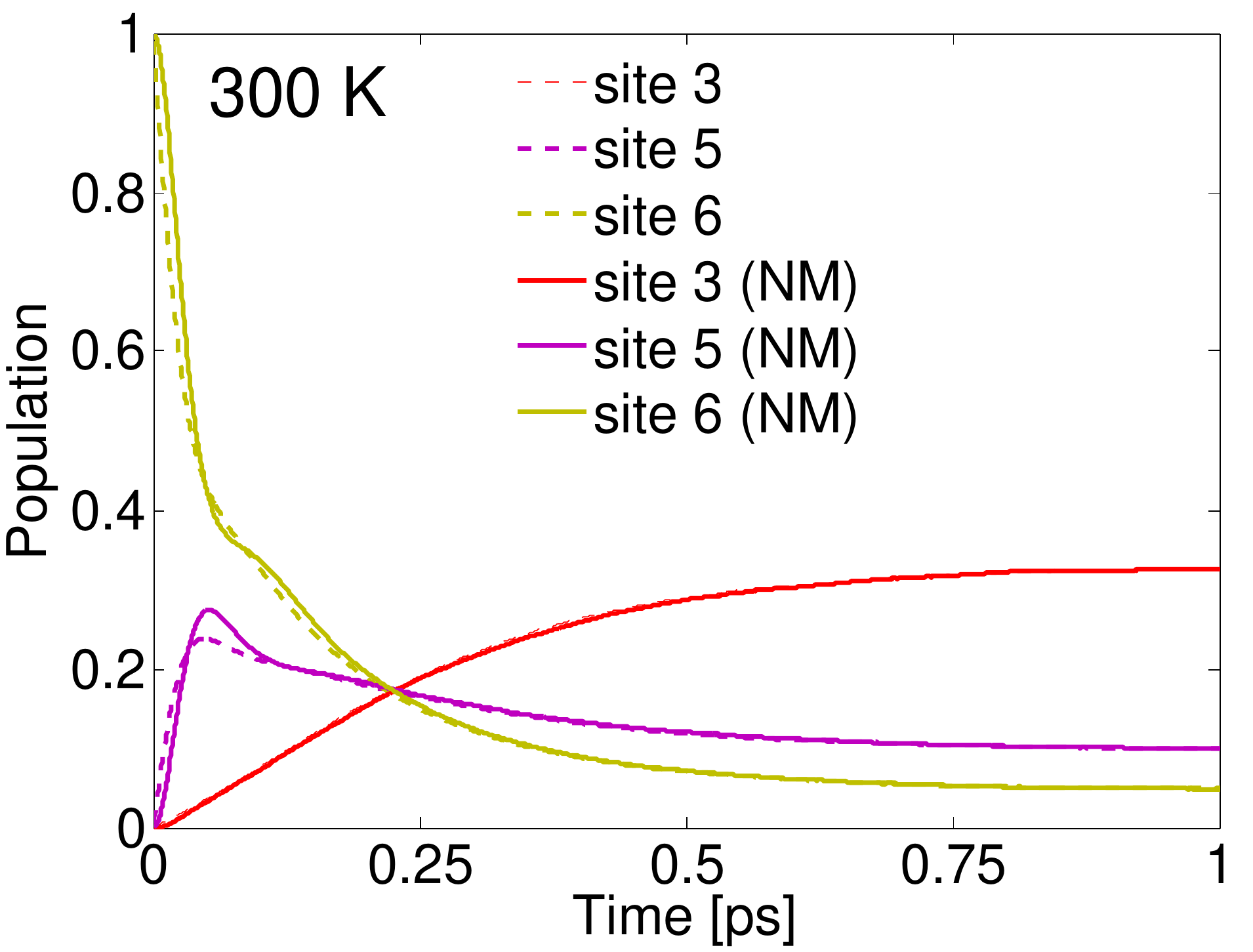}
\caption{Population of sites in the Fenna-Matthews-Olson complex for
Markovian (dashed) and non-Markovian (solid) cases. Sites that are close to
the chlorophyll antenna are taken to be the initial states, i.e.~site 1
(upper panel) or site 6 (lower panel). Parameters are $\protect\omega_c=150$%
cm$^{-1}$, $\protect\lambda=35$cm$^{-1}$, and $T=300$K. }
\label{figureFMO}
\end{figure*}

We also performed simulations for the Fenna-Matthews-Olson (FMO) complex.
The FMO complex acts as an energy transfer wire in green sulphur bacteria
\textit{Chlorobium tepidum} \cite{Fleming09}. It is subject of recent experimental
\cite{Engel07} and theoretical studies \cite%
{Adolphs06,Mohseni08,Ishizaki09_3,Sarovar09}. We derived the TCL master
equation (\ref{eqNonMarkovianMasterEquation}) for the seven-site FMO subunit
and performed simulations with NMQJ method. We used the Hamiltonian of ref.
\cite{Ishizaki09_3} and the spectral density (\ref{eqSpectralDensityOhmic})
with $\lambda =35$cm$^{-1}$ and $\omega _{c}=150$cm$^{-1}$ \cite{Mohseni08}.
The decoherence rates for the 42 relaxation channels (absorption + emission)
and the 7 dephasing channels are evaluated. The rates oscillate and some
have negative regions. The initial states are chosen to be localized at site
1 or 6, the sites that are close to the chlorophyll antenna complex. As a
result, we find that the dynamics is not substantially affected by the
time-dependent rates, see Fig.~\ref{figureFMO}. Quantum beatings are
slightly longer-lived for both temperatures $77$K and $300$K and both
initial states. This is because the NM dephasing rates converge from below
to the Markovian limit similar to Fig.~\ref{figureSpectralDensity} (bottom
right panel). The main relaxation rates stay positive and oscillate around
their Markovian values. The spectral density is rather broad, covering all
transition frequencies, cf.~Fig. 2 of \cite{Adolphs06}, such that the
effects described in the previous sections turn out to be not dominant. The
Markovian approximation alone in the presence of the other approximations
such as Born and secular does not have a substantial effect. Recently,
Ishizaki and Fleming utilized the hierarchical equation of motion approach
for explaining long-lived coherences in the FMO complex \cite{Ishizaki09_3}. Since Born and secular approximations
are avoided for Gaussian fluctuations, this approach has a larger range of validity than the Redfield model,
especially at large temperatures,
and correctly incorporates molecular reorganization effects.

\section{Conclusion}

In conclusion, we have applied the non-Markovian quantum jump method to
excitonic energy transfer. The NM decoherence rates resulting from a
time-convolutionless treatment of the master equation are oscillatory and
negative for parameter regimes and time scales that are relevant to the
problem. In the present work, NM
effects are large when a system is strongly coupled to \textquotedblleft
off-resonant" phonon modes of the environment. These slow modes can lead to
population beatings at room temperature, which are a signature of bath
recurrence effects. Additionally, our computations show that Markovian versus non-Markovian dynamics
can thus crucially affect transport dynamics. Quantum transport can be enhanced over strictly
Markovian dynamics due to a sampling of broader regions of the spectral density.
We thus have provided a non-Markovian extension to recent environment-assisted quantum
transport (ENAQT) concepts. For
example, a system with a transition of around $140$cm$^{-1}$ shows
considerable NM improvement of transport in the presence of strong modes at
around $30$cm$^{-1}$.

Recently, Jang \textit{et al.} \cite{Jang08} developed a novel theory of coherent
resonance energy transfer. A small polaron transformation is applied
before the second-order time-convolutionless expansion that leads to
time-dependent decoherence rates.
Nonequilibrium reorganization effects are
taken into account by the exciton-phonon dressed state description.
This treatment can lead to an increased range of validity with respect to the
exciton-phonon couplings and temperatures compared to the standard Redfield approach.
In this context, the NMQJ method in its present form or with suitable extensions
could prove especially powerful to efficiently simulate larger
donor-acceptor systems and to correctly incorporate negative decoherence
rates in a quantum jump description. Furthermore, the NMQJ method can be used for
the stochastic computation of spectroscopic signals \cite{Sanda08,Mancal06}.

A. A.-G. and P. R. thank the Department of Energy (DOE) M.I.T./Harvard/Brookhaven Center
for Excitonics for support. R. C. thanks the Harvard PRISE program for their support during the
summer of 2009.

\end{document}